\documentclass[sigconf, authorversion]{acmart} %

\usepackage{enumitem}
\usepackage{xspace}
\usepackage{multirow}
\usepackage{backref}

\newcommand{\jump}[1]{\emph{JumpExTra VR}\xspace}

\newcommand{\comm}[1]{}
\newcommand{\compresslist}{
  \setlength{\itemsep}{1pt}
  \setlength{\parskip}{0pt}
  \setlength{\parsep}{0pt}
}

\usepackage{graphicx} 
\usepackage{tikz}
\usetikzlibrary{shapes,arrows}
\usetikzlibrary{positioning,trees
,shapes.geometric, arrows, calc, fit}
\usepackage{xcolor}
\definecolor{mygreen}{RGB}{49,163,84}

\newcommand{\quoting}[2][P]{``\emph{#2}''\emph{(\textbf{#1})}}
\newcommand{\mean}{\emph{M}=}
\newcommand{\sd}{\emph{SD}=}

\newcommand{\psmall}{\emph{p}$<$0.001}

\newcommand{\rev}[1]{#1}

\AtBeginDocument{%
  }

\copyrightyear{2023} 
\acmYear{2023} 
\setcopyright{acmlicensed}\acmConference[CHI '23]{Proceedings of the 2023 CHI Conference on Human Factors in Computing Systems}{April 23--28, 2023}{Hamburg, Germany}
\acmBooktitle{Proceedings of the 2023 CHI Conference on Human Factors in Computing Systems (CHI '23), April 23--28, 2023, Hamburg, Germany}
\acmPrice{15.00}
\acmDOI{10.1145/3544548.3580973}
\acmISBN{978-1-4503-9421-5/23/04}

\begin{document}

\title[Training the Lower Body with Vertical Jumps in a VR Exergame]{Never Skip Leg Day Again: Training the Lower Body with Vertical Jumps in a Virtual Reality Exergame}

\author{Sebastian Cmentowski}
\authornote{Both authors contributed equally to this research.}
\orcid{0000-0003-4555-6187}
\affiliation{%
  \department{High-Performance Computing} \institution{University of Duisburg-Essen}
  \city{Duisburg}
  \country{Germany}}
\email{sebastian.cmentwoski@uni-due.de}

\author{Sukran Karaosmanoglu}
\authornotemark[1]
\orcid{0000-0002-9624-4258}
\affiliation{%
\department{Human-Computer Interaction} 
  \institution{Universität Hamburg}
  \city{Hamburg}
  \country{Germany}
    \streetaddress{Vogt-Kölln-Straße 30}
  \postcode{22527}}
  
\email{sukran.karaosmanoglu@uni-hamburg.de}

\author{Lennart E. Nacke}
\orcid{0000-0003-4290-8829}
\affiliation{
\department{HCI Games Group \& Stratford School of Interaction Design and Business} 
\institution{University of Waterloo}
    \city{Waterloo}
  \country{Canada}}
\email{lennart.nacke@acm.org}

\author{Frank Steinicke}
 \orcid{0000-0001-9879-7414}
\affiliation{
\department{Human-Computer Interaction} 
  \institution{Universität Hamburg}
 \city{Hamburg}
  \country{Germany}
    \streetaddress{Vogt-Kölln-Straße 30}
  \postcode{22527}}
 \email{frank.steinicke@uni-hamburg.de}
 
\author{Jens Krüger}
\orcid{0000-0002-9197-0613}
\affiliation{%
\department{High-Performance Computing}
  \institution{University of Duisburg-Essen}
  \city{Duisburg}
  \country{Germany}}
\email{jens.krueger@uni-due.de}

\renewcommand{\shortauthors}{Cmentowski and Karaosmanoglu et al.}

\begin{abstract}
Virtual Reality (VR) exergames can increase engagement in and motivation for physical activities.
Most VR exergames focus on the upper body because many VR setups only track the users' heads and hands.
To become a serious alternative to existing exercise programs, VR exergames must provide a balanced workout and train the lower limbs, too. To address this issue, we built a VR exergame focused on vertical jump training to explore full-body exercise applications. 
To create a safe and effective training, nine domain experts participated in our prototype design. 
Our mixed-methods study confirms that the jump-centered exercises provided a worthy challenge and positive player experience, indicating long-term retention. 
Based on our findings, we present five design implications to guide future work: avoid an unintended forward drift, consider technical constraints, address safety concerns in full-body VR exergames, incorporate rhythmic elements with fluent movement patterns, adapt difficulty to players' fitness progression status.

\end{abstract}




\begin{teaserfigure}
  \includegraphics[width=\textwidth]{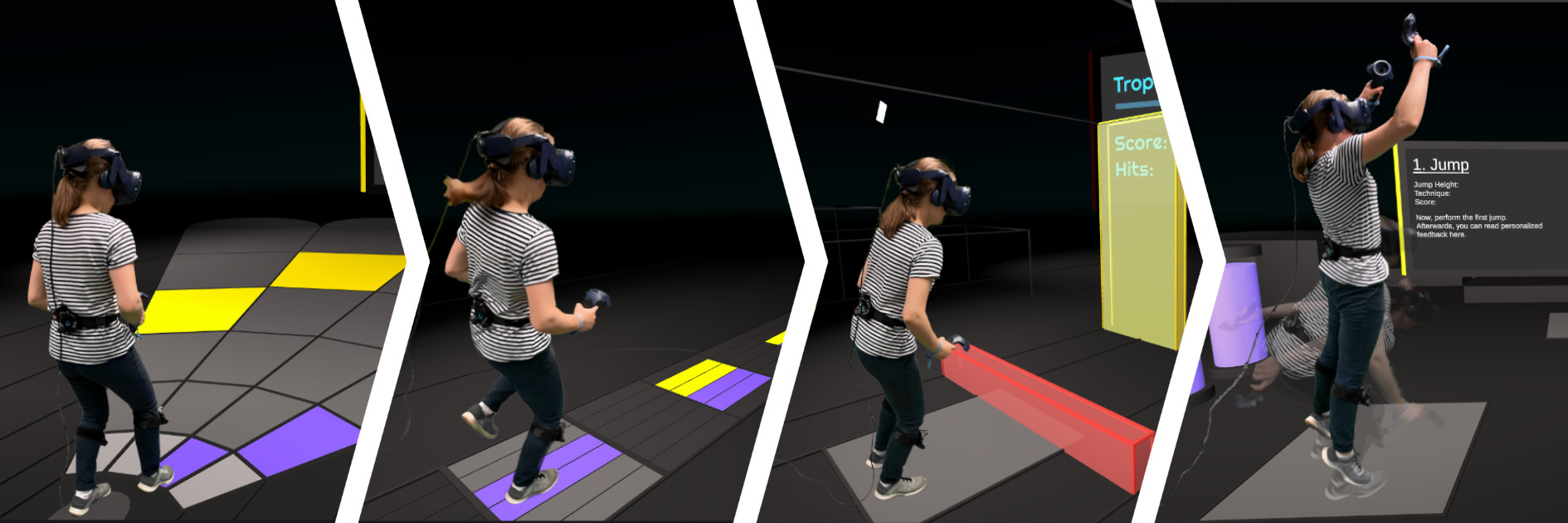}
  \caption{Demonstration of a VR-based full-body training with our exergame \jump{}: In the first three levels (from left to right), players perform various movements, including taps, hops, and jumps, that train lower body coordination, stability, and endurance. Finally, the right figure shows a player who trains maximal vertical jumps.}
  \Description{The figure has a total of four sub-figures, which shows the sequential levels of \emph{JumpExTra VR}. From left to right: the player taps with the right foot on a colored tile, the player jumps with one foot on the spot, the player prepares to jump over an approaching hurdle, and the player performs a maximal vertical jump with an arm swing.}
  \label{fig:teaser}
\end{teaserfigure}

\maketitle

\section{Introduction}

Regular physical exercise is vital for our bodily and mental well-being. Athletic activities not only increase our overall fitness but can also delay the natural aging process~\cite{erickson2011exercise} and even benefit the brain's cognitive functions~\cite{nouchi2014four}. However---because physical activity is strenuous on our bodies---many people hesitate to transition toward an active lifestyle and to create lasting exercise habits if they do not receive incentives~\cite{charness2009incentives,gabel2021corona}. Apart from popular approaches, such as peer-support or fitness trackers, exergames\rev{---\emph{``digital game[s] where the outcome [...] is predominantly determined by physical effort''}~\cite{muller2011designing}---}promise to motivate users by providing an engaging experience. In the virtual reality (VR) domain, fitness games are among the highest-grossing titles and a crucial reason for headset purchases~\cite{gallagher_2022}. The advantages are apparent: Affordable mobile VR headsets with handheld controllers allow players to combine enjoyable gaming activities with healthy physical exercise while staying in the comfort of their homes.

Exergames like Beat Saber\footnote{\rev{We note that Beat Saber is a VR rhythm game. However, some users can use this game for exercise purposes, and this game has been the subject of research as an exergame~\cite{szpak2020exergaming,albert2022effect}.}}~\cite{beatsaber} or \rev{FitXR~\cite{fitxr}} mainly use \rev{head and hand movements for their gameplay while featuring some lower-body movements. Although there are also VR applications that target full-body exercises (e.g., OhShape~\cite{ohshape} and VRWorkout~\cite{VRWorkout}), the majority only use lower body movements indirectly, such as when ducking under obstacles. Such exercises are easily possible with the default VR setup, i.e., headset and controller tracking, without requiring additional hardware, such as Vive trackers. Consequently, the training effect of these exergames mainly targets} cardiovascular improvements and upper body fitness. As a result, players do not get the full benefits of balanced full-body activities, such as improved coordination, stability, and balance~\cite{anderson2005impact, oliveira2017balance, stovsic2020effects}. Also, training only individual muscle groups can ultimately lead to muscular imbalances promoting bad posture~\cite{kim2006influences, buchtelova2013influence} and increasing the risk for injuries~\cite{sugiura2008strength, croisier2004muscular}. Lastly, lower body exercises and physical activities like walking, running, or jumping are vitally important in our society because an average person spends most of their day sitting and moving insufficiently~\cite{whoPhysicalActivity}. To make VR exergames more valuable as exercise environments, their traditional hand-focused gameplay must be adapted to incorporate activities for the lower body. \rev{While we currently do not know if VR exergames are as effective as gym classes or personal training, we know from short-time studies that exergames provide great motivation for exercising~\cite{macvean2013,martinniedecken2020hiit,finkelstein2011astrojumper,born2019exergaming}. However, we note that long-term exergame studies present mixed results on users' adherence~\cite{robertson2018savouring,uzor2014investigating}.}

Unfortunately, using lower body or full-body movement in VR exergames introduces many challenges. These exercises have a higher risk of swift or unstable movements which might lead to dangerous collisions. Explosive movements, such as running or jumping, can suffer from poor tracking stability~\cite{yoo2016vrun,ioannou2019virtual}, which negatively affects player experience. Full-body movements require expert knowledge for safe training because they are more complex than simple arm swings in Beat Saber~\cite{beatsaber} and bear a greater risk of injury or wrong execution. Limitingly, existing design guidelines and best practices primarily target physical exercises and movements for non-VR exergames~\cite{mueller2014, marquez2013, mueller2016}. %

Our research fills this knowledge gap by following the feedback from domain experts to create a full-body VR exergame. We focus on one particular use case: training people's vertical jump performance using a VR exergame. We chose the vertical jump exercise as our full-body movement because of the following reasons: Jumping is a fundamental human movement that is not only required for many sports, such as basketball~\cite{decker1996using} or volleyball~\cite{powers1996vertical}, but is also used to assess general fitness~\cite{graham1994guidelines}, body composition~\cite{bosco_mechanical_1983}, and functional performance~\cite{mcmaster_brief_2014}. Vertical jumps are a perfect, yet challenging, core movement for our research. They also work inside the tracking areas of current VR headsets. At the same time, vertical jumps are a highly explosive movement that challenges tracking stability. Finally, jumps can be improved through many training modalities and combined with other movements to achieve a diverse exercise experience.

We focus primarily on improving general fitness and motivation through training. For this, we conducted a semi-structured interview with experts from different domains (e.g., sports research or physical therapy). We discuss the potential benefits and challenges of VR-based jump training identified in our thematic analysis and provide guidelines for structuring gamified training routines. We developed a VR exergame to train the vertical jump based on these insights. In our design process, we closely follow the recommendations of experts. In particular, our exergame is composed of four levels of increasing difficulty to prevent injuries and foster the learning process (see~\autoref{fig:teaser}).

\rev{In the second phase of our work}, we investigate how users perceive our exergame prototype and what implications can be drawn for future designs and research projects. As our first step, we conducted an exploratory study with 25 participants to evaluate how users perceive this new training experience. The results confirm that our jump-centered exercises provided a worthy challenge and led to a positive player experience. Our study also revealed the technical limitations of current VR systems, and the participants provided substantial suggestions for improving the training experience. Subsequently, we condensed these insights into design implications and lessons learned.

\rev{In the design implications, we first} discuss potential safety issues. In our case, frequent jumps on one spot often led to an unnoticed forward movement---the unintended forward drift---which eventually leads to players leaving the intended play area. Hence, we recommend particular care to avoid dangerous collisions, especially when non-stationary movements, such as forward jumps, are used. %

Next, we examine the design implications arising from technical limitations of current VR systems. Slipping hardware trackers and insufficient tracking accuracy challenged the precision of our individualized jump feedback. Therefore, we recommend empowering users towards self-correction (e.g., by visualizing a replay of their movements) and supporting this process with automated feedback. 

Finally, we talk about \rev{the design implications based} our efforts to provide a pleasant game experience and increase replayability. Above all, the participants praised \rev{using our exergame} \jump{} \rev{for jump training}. %
We discuss lessons learned for enabling a natural and fluent movement sequence, such as the suitability of different patterns (e.g., the ``walking style'') or incorporating frequent resting periods. %
We emphasize the importance of aligning the difficulty with the players' capabilities and improvements. In particular, beginners profit from adapting the difficulty automatically.

The main contributions of this research are:
\begin{enumerate}[noitemsep,topsep=0pt]
\item Identifying the benefits and requirements of VR jump training through semi-structured interviews with domain experts,
\item Designing and developing a VR exergame prototype for training the vertical jump: \jump{},
\item Conducting an exploratory study to evaluate the feasibility of our exergame as a training tool, and
\item Deriving a set of design guidelines and lessons learned for developing full-body VR exergames.
\end{enumerate}

Our explorative exergame study and the resulting five design guidelines (unintended forward drift fix, technical constraints consideration, safety concern mitigation, rhythmic elements using fluent movement patterns, difficulty adaptation to players’ fitness level) constitute a first step to the creation of safer and more engaging exergame VR training environments. Our findings help players train effectively and without the risk of injury in small spaces using immersive technology. We believe this represents one possible future of technology-augmented sports exercises.

\section{Related Work}

Our exergame design builds on two domain knowledge sources: prior research and expert interviews. Therefore, we introduce VR-based training, the relevant biomechanical foundations of the vertical jump and provide an overview of the related research on jump training, jumping in VR, and exergames in general.

\subsection{VR Training}
Using immersive experiences for training offers unique benefits and allows users to assess and improve their individual performance effectively~\cite{cannavo2018movement, van2009virtual}. Especially when physical training at the target location is difficult or dangerous, like for pilots~\cite{hays1992flight} or firefighters~\cite{stansfield2000design}, VR applications can provide an accessible alternative~\cite{jerald2015vr}. In recent years, VR-based training has been successfully applied to many domains, including healthcare~\cite{mantovani2003virtual}, medicine~\cite{al2006using}, and navigation~\cite{mas2018indy}. In sports, VR has been used in research projects to analyze athletic performances~\cite{bideau2010virtual, craig2013understanding} and understand motor and perceptual skills~\cite{craig2006judging, chardenon2002visual, fink2009catching, zaal2011virtual}. Furthermore, VR works well for training movement patterns, like golf swings~\cite{kelly2010virtual} or dance moves~\cite{eaves2011short}, and improving hand-eye coordination (e.g., for table tennis~\cite{michalski2019getting}, darts~\cite{tirp2015virtual}, or juggling~\cite{lammfromm2011transfer}). Similarly, VR training can also support rehabilitation for stroke or cerebral palsy patients~\cite{chen2007use, mirelman2010effects, kim2009use}. %

Besides boosting motivation, virtual experiences also have benefits for sports training. VR improves observational learning by displaying correct action patterns immersively compared to traditional computer applications~\cite{tanaka20173d}. Especially for sports, stereoscopic information is essential to trigger the correct motor responses~\cite{yeh1992spatial, lenoir1999ecological}. Additionally, mobile headsets also provide an easy way of monitoring the own performance~\cite{neumann2018systematic}. For availability, VR applications can---to some degree---eliminate the need for specialized sporting equipment, dedicated training environments, or workout partners~\cite{michalski2019getting}. Adaptive training routines~\cite{duking2018potential} and personalized feedback increase the individual gain and can complement professional trainers~\cite{kim2013unsupervised}. Despite these benefits, the effectiveness of VR applications for sports training remains an open research field. While studies have shown a positive impact on the execution of a particular exercise, the transferability to actual activities is not guaranteed~\cite{lathan2002using, baldwin1988transfer}. VR training can improve real-world skills for some use cases~\cite{michalski2019using, bliss1997effectiveness, carlson2015virtual}, but others cannot profit in the same way~\cite{kozak1993transfer} and might even suffer from reduced performance~\cite{todorov1997augmented}.

\subsection{Jumping}

Jumping is a fundamental motor skill humans learn at an early age and improve upon throughout their lives. While most people do not jump regularly in their day-to-day lives, jumping is used in sports, fitness, and rehabilitation. In particular, jumps are a good indicator of a person's general fitness level~\cite{graham1994guidelines}, functional performance~\cite{mcmaster_brief_2014}, and muscle composition~\cite{bosco_mechanical_1983}. Apart from being required for typical jumping-intensive sports, such as basketball~\cite{decker1996using},%
jumps are used in rehabilitation to measure changes in pathology~\cite{hegedus2015clinician, helland_mechanical_2013} and predict the individual risk for injuries~\cite{cesar_frontal_2016}. Aside from the positive effects on athletic performance~\cite{canavan2004evaluation, potteiger1999muscle}, jump training can also benefit daily activities and occupational tasks~\cite{kraemer2001effect, bassey1992leg}.

\subsubsection{Composition and Execution of the Vertical Jump}
Jumping is a ``complex polyarticular dynamic movement requiring intermuscular coordination''~\cite{perez2013training} with typical execution times of less than 4 seconds~\cite{vanezis2005biomechanical}. %
The literature differentiates between various types of jumps~\cite{bobbert1990drop}, including squat jumps, drop jumps~\cite{hunter2002effects}, or \textit{countermovement jumps}. In the scope of this paper, we focus on the last type. It is typically initiated from an upright standing position~\cite{burkett2005best}, followed by a brief downward phase before the upward impulse is generated by extending the body explosively and swinging the arms in a forward-upward arc~\cite{hough2009effects}.
        
Jumping can exert high loads on the lower body's joints and tissues, increasing the risk of injury %
(e.g., ruptures of the anterior cruciate ligaments (ACL)~\cite{piasecki2003intraarticular} %
caused by hard or incorrect landings that lead to high vertical ground reaction forces (vGRF)~\cite{hewett2005biomechanical}). %
Proper instructions can help athletes reduce the vGRF immediately~\cite{mcnair2000decreasing,prapavessis1999effects}. Other risk factors, such as the athletes' joint stiffness~\cite{devita1992effect} and maturity~\cite{lazaridis2010neuromuscular}, cannot be eliminated as easily with proper form. 

\subsubsection{Jump Tests}
Measuring a person's jump height is not a trivial task. Professional athletes and researchers often rely on motion-capturing systems~\cite{balster2016effects, bates2013impact, lees2004maximal} or dedicated vertical jump tests~\cite{klavora2000vertical}. One of the oldest techniques is to jump next to a wall and mark the highest reachable points while standing and jumping. Then, one can calculate the effective jump height from the difference between both points~\cite{jensen1972scientific}. Whereas this jump and reach test follows a simple principle, it also splits the athlete's attention and limits arm movement, which easily reduces performance. Another approach is to jump on force platforms and calculate the jump height from the athlete's airtime~\cite{bosco_mechanical_1983}. However, athletes can easily distort the result by flexing their legs to delay ground contact~\cite{klavora2000vertical}. Instead, the most precise results are achieved by measuring the vertical displacement of the athlete's center of mass~\cite{vanezis2005biomechanical}. Consequently, our VR application can easily and comfortably measure the precise jump height by using a hardware tracker attached to the user's hip.

\subsubsection{Jump Training}
Critical for improving jump height is increasing an athlete's take-off velocity. Training usually concentrates on improving the extensor muscles' forces and contraction velocity~\cite{kyrolainen2004effects} because the maximal dynamic force of the lower extremities and the rate of force development (RFD) directly correspond to the final jump height~\cite{blackburn1998relationship, wilson1996use}. %
A well-timed arm swing~\cite{carlson2009effect}, good neuromuscular control to initiate joint extension with minimal delay~\cite{hudson1986coordination}, and a countermovement with the optimal squat depth~\cite{gheller2014effect} have all been shown to improve jump height significantly.

Various training modalities have been found useful in achieving lasting training results. Firstly, plyometric exercises~\cite{adams1992effect}, such as drop jumps or alternate-leg bounding~\cite{fleck2014designing}, improve the muscles' stretch-shortening cycle (SSC) and are primarily beneficial activities that involve rapid concentric contractions and high-intensity eccentric contractions~\cite{malisoux2006stretch}. The positive effect on jump performance has been extensively researched~\cite{bobbert1990drop, lundin1991plyometrics} and varies between 5\% and 35\% depending on the athletes' proficiency~\cite{perez2013training}. Alternatively, athletes may conduct weight training~\cite{wilson1996weight} with heavy loads to increase the maximal dynamic strength~\cite{gabriel2006neural} or light loads for explosive movements~\cite{lyttle1996enhancing}. For jump training, the best results were reported using light loads and high speeds and range from 2\% to 25\%~\cite{perez2013training}.
Some muscles, such as the hip joint extensors, are used only maximally in maximal jumps.%
Performance should be trained using maximal jumps to achieve the best training results~\cite{lees2004maximal}.

\subsection{Jumping in VR and Exergames}
Jumping in virtual and mixed reality has also been the subject of other research projects. 
Prior literature has approached jumping in the context of locomotion~\cite{wolf2020jumpvr,hayashi2019redirected,li2021detection}, exergames~\cite{finkelstein2010astroautism,ioannou2019virtual,kajastila2014empowering,lehtonen2019movement} or training~\cite{cmentowski2021exploring}.
\citet{wolf2020jumpvr} presented an augmented locomotion technique where users performed physical vertical jumps, which translated into hyper-realistic forward jumps in VR. Their findings indicated that hyper-realistic jumps can enhance some factors of user experience (e.g., immersion). 
A recent work~\cite{cmentowski2021exploring} presented a prototype of a VR jump training where players jumped and received feedback on their performance.%

VR exergames are one of the most popular types of games in the VR gaming community (e.g., Beat Saber~\cite{beatsaber}) \rev{and have been shown to elicit a higher level of motivation compared to their non-VR headset exergame counterparts~\cite{born2019exergaming}.}
Despite the potential drawbacks associated with the use of VR exergames (e.g., cybersickness~\cite{szpak2020exergaming}), we see these games as an opportunity because they offer fun, physical activity, and accessibility to training regardless of location or health condition~\rev{\cite{karaosmanoglu2021lessons,kruse2021longterm}}. 
\rev{Although prior research also covered exergames featuring full-body training~\cite{martin2019exercube,kajastila2014empowering,ioannou2019virtual} and investigated full-body movement recognition~\cite{caserman2022fullbody}, many papers focus specifically on upper-body exercises~\cite{karaosmanoglu2021lessons,born2021motivating,karaosmanoglu2022canoevr}}.
Whereas this focus may be even preferable in some cases (e.g., due to safety~\cite{karaosmanoglu2021lessons}), full-body training could benefit more muscle structures, and has not been widely explored yet~\cite{martin2019exercube}. 

\rev{Many papers explored the design space for exergames and provided implications for designers, developers, and researchers~\cite{mueller2014,mueller2016,marquez2013}. However, to our knowledge, these design guidelines have not been specifically focused on VR exergames.
\citet{marquez2013} considered technological, social, and physical factors in designing \emph{``body games''}. In their approach, they primarily focus on understanding the social and physical factors around a game and accordingly support the users with the technology.
Furthermore, \citet{mueller2014} provided ten comprehensive guidelines for movement-based games, such as (i) using tracking inaccuracies, (ii) rhythmic elements, and (iii) utilizing risk reasonably as elements in movement-based gameplay.
Similarly, \citet{mueller2016} presented five design recommendations for exertion games, or exergames: (i) providing an easy start, (ii) presenting short-term achievable goals for long-term motivation, (iii) considering individual skill levels of players, (iv) giving feedback on the performance, and (v) employing social play to promote motivation.
Whereas these guidelines are helpful for movement-based games in general, full-body VR exergames may present additional advantages (e.g., higher motivation in VR exergames~\cite{born2019exergaming}) or challenges (e.g., wearing a headset) that require attention.}

Several researchers have used jumping in their games. 
\citet{finkelstein2010astroautism} designed an CAVE-based exergame, \emph{AstroJumper}, for children with autism.
In the game, the players performed jumping movements to avoid objects, and the initial findings with neuro-typical players indicated positive experiences. 
\citet{kajastila2014empowering} examined the impact of three conditions on players' learning trampoline skills. 
The authors showed that the players were more engaged in the gaming conditions (a trampoline-based mixed reality game with and without exaggerated jumps) compared to the self-training condition, but their performance improved regardless of the conditions. 
Similarly, another study~\cite{lehtonen2019movement} designed and tested a multiplayer mixed reality trampoline game in a field study.
The results indicated positive player experiences (e.g., autonomy and physical activity enjoyment). 
Many jump-based exergames have focused on %
player experience rather than providing a structured physical training for jumping using VR exergames. 
This paper extends the previous work~\cite{cmentowski2021exploring} by designing and testing a detailed jump training VR exergame, \jump{}, and involving the domain experts in the process.

\section{Thematic Analysis of Expert Interviews}
\begin{figure*}[ht]
\pgfdeclarelayer{background}
\pgfsetlayers{background,main}
\definecolor{expert1}{HTML}{54278f} 
\definecolor{require1}{HTML}{807dba}
\definecolor{imp1}{HTML}{bcbddc} 
\definecolor{user1}{HTML}{fed976} 
\definecolor{pre1}{HTML}{ffeda0}
\definecolor{study1}{HTML}{ffeda0}
\definecolor{post1}{HTML}{ffffcc}
{
\normalsize
\resizebox{\textwidth}{!}{
\includegraphics{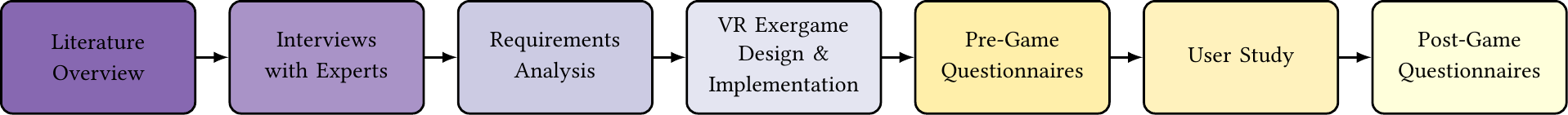}
}}
\caption{The design process followed in this research project: We reviewed the literature, conducted semi-structured interviews with multiple experts, analyzed the requirements of our VR jump training application, designed and implemented the VR exergame, and conducted a user study to test \jump{}.}\label{fig:procedure}
\Description{The figure shows a diagram that illustrates the design process followed in this research project. From left to right, the diagram lists the following steps: literature overview, interviews with experts, requirement analysis, VR exergame design \& implementation, pre-game questionnaires, user study, and post-game questionnaires.}
\end{figure*}

Unlike the prior literature, our research was not only on professional athletes and their sports performance training but we focus on the motivation of players to exercise.
We gathered information from experts in the sports and medical field through semi-structured interviews to design and implement \jump{} (see~\autoref{fig:procedure}).
After internal discussion, the first authors created the interview questions and selected areas of expertise for interviewees, such as sports physicians and trainers.
Based on these decisions, they selected several experts and invited them to this study via email.
Nine experts from different domains (i.e., sports research, physiotherapy, and training) were recruited for the interviews.
Given the variety of domains, the first authors roughly followed the initial interview guidelines (listed in the supplemental materials), but occasionally deviated from these questions to account for the experts' specialties.
The interviews were conducted using a video conference tool in German and took 42.4 minutes on average.

To prepare the interview data for analysis, we used the Dovetail~\cite{dovetail} software for transcription.
One of the first authors checked these transcriptions to correct and cut unnecessary details.
Then, these texts were translated into English using DeepL Pro~\cite{deepl} and then checked again for any errors.
To analyze the interview data, we used characteristics of both reflexive and codebook approaches of thematic analysis~\cite{braun2021can,braun2019reflecting}.
Before the analysis, the first authors identified deductive categories related to the research questions: \emph{execution}, \emph{importance of jumping}, \emph{safety in jumping}, and \emph{jump training}.
Then, these two authors independently and inductively coded the interview in groups, using descriptive codes (e.g., ``correct execution of a counter-movement jump with arms'', ``landing is a major source for injuries'', and ``few short term improvements'') under these categories.
The groups consisted of two interviews in each and three interviews in the last one.
After each group, the first authors met to discuss discrepancies between their coding and each author's understanding of the data.
This process led to the creation, refinement, combination or exclusion of codes.
Following the last group of interviews, the first two authors developed initial themes by creating a affinity map from those codes.
Finally, they discussed and reshaped the affinity map and these themes in several meetings, which led to the formation of the following four themes. 

\subsection{Theme 1: Jumps are mainly used in sports and have numerous benefits for general health.}

Both in everyday life and sports activities, we \textit{jump}.
However, the actual form depends on the particular context, as 
\quoting[E3]{[...] we don't jump that often when we go shopping in the supermarket, we probably walk more. But of course you also have situations in which you might jump down in everyday life}.
Instead, jumps are mostly \textbf{integrated into compound movements}.
For example, they might be combined with forward or side movements to dodge obstacles.
Especially in sports, jumps are mostly incorporated into bigger movement patterns, such as block jumps in volleyball.

We perform jumps because they have several benefits for general health.
The jump movement activates multiple muscle structures and allows us to exercise all of them at once.
Moreover, it can help to improve fundamental human skills like \textbf{coordination}.
A good execution requires the jumper to \quoting[E2]{[...] coordinate the impulse from the arms with the impulse on the legs [...]}.
A physiotherapist emphasized the importance of jumps for \textbf{injury prevention} and gave an example of why it can help us: \quoting[E6]{If I am an untrained person and I trip over a curb I could twist my ankle. But if I have trained [to jump] and [thereby] manage to activate my muscles quickly, I might be able to stabilize my body in time so that this accident doesn't happen}. 
Despite this potential for injury prevention, jump training rarely finds applications in rehabilitation because health insurance often covers \quoting[E6]{only the bare necessities}.
By jumping, the risk of some physical health conditions, such as \quoting[E2]{osteoporosis}, can be decreased.
Interestingly, jumping, in fact, might be a helpful tool to trigger the regrowth of bones for older adults with reduced bone density. 
However, it might also be dangerous as \quoting[E5]{maybe they don't have the stability yet to catch themselves and break away}.
Therefore, most experts indicated that jump training is generally not favorable for older adults. 
In particular, one expert pointed out that alternatives should be considered to reach similar benefits: \quoting[E2]{Do jumps then make much sense or don't you achieve that rather, for example, with more strength training equipment?}.

In general, we mainly use jumps for \textbf{testing, training, and in sports games}.
First, jumps are a good instrument to test people's performance, and can lay down a foundation to understand \quoting[E1]{[...] the rate of force development, i.e., how quickly can I generate force, which is well represented by jumps}.
However, the primary application of jumps lies in the actual gameplay.
Especially in certain sports such as basketball and volleyball, jumping can be a decisive part of the game.
For instance, an interviewed basketball trainer noted that a good jump performance could be a game-changing advantage for some players: \quoting[E7]{Jumping power itself is particularly relevant in basketball because you can compensate for your physical size a bit, if necessary}.
Finally, using jumps in training, e.g., for volleyball, can be beneficial to improve the fatigue capacity: \quoting[E1]{Someone who practices a lot and is well trained in jumping can hold the jump height much longer with a lot of jumps before it gets less}.

\subsection{Theme 2: Jump training combines reactive and strength exercises. Incorrect executions of maximal jumps leads to injuries.}

Domain experts frequently gave the obvious answer to how to best train for better jumps: \quoting[E1]{by jumping}. However, apart from this conspicious statement, it is essential to frame individual exercise goals: increasing the jump power requires different training concepts from working on the jump technique.

Jump power is mainly determined by the muscles' rate of force development. Hence, it can be improved through various training modalities. In particular, \textbf{explosive movements} and \textbf{speed-based exercises} are especially effective. An interviewed basketball trainer described a typical reactive training they regularly performs: \quoting[E7]{We put several [boxes] in a row behind each other and did bounce training there with our legs closed. Always both feet on a box and then further, up, down, up, down and so moved through the hall}. Additionally, exercises that target the muscles' lift capabilities, e.g., traditional strength training, also improve the overall jump height. Also, the experts recommended \textbf{combining jump training with other physical exercises} and including \quoting[E2]{different variations} as jumps are usually not performed in isolation. Unfortunately, short-term improvements are not to be expected as changes in muscle mass and composition typically take many weeks: \quoting[E5]{It usually takes 4 to 8 weeks minimum, until you can really prove something muscular or microscopically}.

Even though small hops are generally unproblematic, most experts agreed that the maximal vertical jump is a highly complex and error-prone movement. Thus, the foremost goal of maximal jump training should be \textbf{correct execution}. Given the variety of used jump techniques, we explicitly asked the experts about the correct execution of a countermovement jump with an arm swing. This movement is typically executed by starting in an upright shoulder-wide stance. After a quick downward movement, the athlete extends their body explosively. The arms swing in an upward arc and are stopped roughly at chest height before the feet lose contact with the ground: \quoting[E2]{If I want to jump to the maximum, I actually have to slow the arms down at shoulder height and at the right moment}. Despite this general routine, the individual body shape and the sports context can lead to differences in the execution, e.g., athletes cannot reach lower squat depths if the gluteal muscles are too weak.
	
Most experts deemed \textbf{landing} after a jump the major source of \textbf{injuries}. A particular pain point is the knee movement. The knees should always remain in one line between the ankle and the hip. However, high-impact forces can cause the knees to collapse medially. This knock-knee position is perilous and puts extreme pressure on the knee and the surrounding ligaments. Apart from a weakly developed musculature, gender differences contribute to this condition as females are generally more prone to having knock-knees: \quoting[E2]{There are also anatomical reasons, it's a little bit due to the hip position of women}. Lastly, preinjuries can increase the risk of further accidents while jumping. Apart from an incorrect landing, the experts mainly attributed the \textbf{exercise environment} as an important injury factor and suggested using a mat as a protective measure. In contrast to the muscular changes, improvements in jump execution are quickly achievable but hard to quantify. One expert proposed measuring the knee deviation during landing as a possible improvement.

\subsection{Theme 3: Jump training should increase gradually and account for individual differences, goals, and improvements.}

In contrast to general physical activity, training is always \textbf{goal-oriented} and consists of exercises that exert a sufficient stimulus to trigger progression, such as muscle growth. However, our interviewed experts underlined the importance of carefully weighing training intensity, repetition count, and recovery time to maximize improvements and avoid injuries. Special attention should be placed on beginners who are not regularly exercising. For instance, \quoting[E5]{if they are not used to it, [their knees and ankles] are very susceptible to evasive movements}.

As the risk of injury depends primarily on the range of motion, training should \textbf{start with simple exercises}, such as mini hops, before gradually increasing the difficulty. Doing so also has another advantage: many experts agreed that small jumps are generally safe and do not require a prior warmup. In contrast, intensive or longer training sessions should be preceded with warmup movements, such as small hops, to prevent muscle strains. One expert whose research focuses on people with special needs proposed to even start with simple steps to make the application accessible for users with coordination and balance problems: \quoting[E9]{So before it's even about doing a jump. To first step over an obstacle, sometimes with the right, sometimes with the left, in order to promote balance}.

Apart from \textbf{adapting} the \textbf{training difficulty} according to the users' abilities and fitness level, experts also emphasized the importance of \textbf{providing proper feedback}. Guiding the users and building competence is vital for sustainable training results. One interviewed sports didact reported the benefits of recording the users' movements \quoting[E3]{so that [they] can see their own jump to create a movement image of themselves}. Combining this intrinsic learning with extrinsic feedback is particularly useful for continuous improvements. Additionally, the experts recommended focusing primarily on repetitive situations that permit users to incorporate their insights in the subsequent execution.

\subsection{Theme 4: VR jump training can provide real-time feedback and boost motivation. Yet, safety might be an issue with VR.}

In general, all experts saw potential benefits and challenges in using VR for jump training. Firstly, a major advantage of VR and AR is the ability to provide directly applicable \textbf{real-time feedback}. Also, whereas some experts recommended using mirrors or video recordings to show the users their movements, many people do not like seeing themselves. In this case, seeing a replay of their own jump might be even counterproductive: \quoting[E3]{Of course, it's not helpful at all to then replay a video of the own moves that aren't working out}. VR can help avoid such potential alienizing effects by introducing an additional abstraction layer. For example, users could see a generic avatar performing a replay of their jump.

Furthermore, VR exergames could benefit mental well-being by improving mood, reducing stress, and increasing the users' \textbf{motivation to train}. In particular, one expert suggested that gamified exercising might provide similar strong incentives like peer support and outperform wearables, as \quoting[E4]{just a fitness tracker [...] can only change something, if so, in the very short term}. However, as we have seen with popular augmented reality (AR) exergames, such as Pokemon Go~\cite{pokemongo}, the users' interest can decrease over time.

Finally, the experts raised concerns regarding the \textbf{safety} of VR-based jump training. For instance, a mismatch between the feedback from the virtual world and the physical movement likely causes cybersickness and could even lead to dangerous situations. Therefore, it is necessary to consider potential issues early in the design pipeline because \quoting[E1]{if [...] the risk of falling or somehow feeling unwell is greater than the benefit I generate, then it's immediately a problem}. Lastly, one expert expressed doubts about whether VR should be used to measure the jump height, as there are likely more affordable and precise approaches.

\section{Exergame: JumpExTra VR}

\begin{figure*}[!ht]
    \centering
    \includegraphics[width=\textwidth]{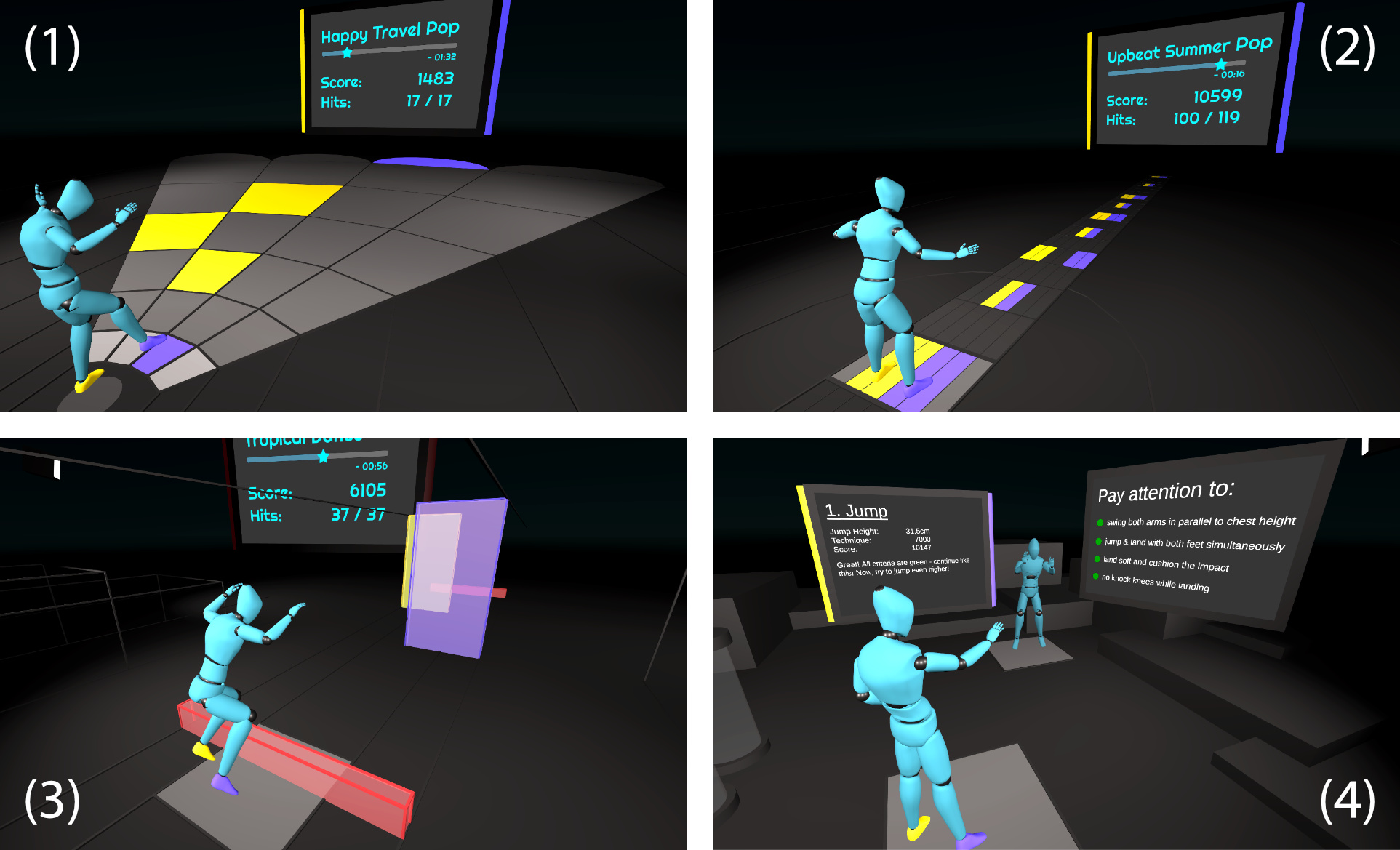}
    \caption{Our exergame \jump{} features four sequential levels. In the first level, players tap on colored tiles with their feet. The second level features a hopscotch game. In the third level, players avoid obstacles. In the last level, players train their maximal vertical jump and receive personalized feedback.}
    \Description{The figure has four subfigures. It provides sequential impressions of \emph{JumpExTra VR}'s game levels and how players perform their movement at these levels. In the first level, players tap on the colored tiles with their feet. The second level features a hopscotch game in which players must jump on their left, right, or both feet. In the third level, players avoid obstacles by performing jumps or side movement. In the last level, players train their maximum vertical jumps.}
    \label{fig:game}
\end{figure*}

Based on the insights from our expert interviews, we designed a VR exergame with the Unity game engine~\cite{unity} to train the vertical jump. Our primary focus was on motivating players to exercise in a gameful way (i.e., using the motivational pull of games). However, we also wanted to avoid any injuries \rev{(i.e., injuries that can be seen during the gameplay, such as hurting a body part due to falling)} or frustration by aligning the gameplay with the players' capabilities and giving feedback to assist them in improving their jump.

The first challenge for implementing a jumping-based exergame is tracking the entire body's movements because most VR systems use only an HMD and tracked controllers. We experimented with various approaches throughout our design process, such as using a Kinect 2 for Windows~\cite{kinect}. However, the high latency and inferior tracking quality when jumping made markerless motion capturing an undesirable choice for our use case. Instead, we opted for using a Vive Pro~\cite{vive} and attaching Vive Trackers~\cite{viveTracker} to the players' shins and waist. Even though tracking players' feet was our initial first choice, our pilot tests revealed that placing the trackers on the shins drastically improves tracking accuracy and comfortability. Together with the controllers and headset, we used a total of six tracking points to animate a virtual avatar using inverse kinematics.  

\subsection{Rhythmic Levels}

Many experts emphasized the importance of raising the difficulty gradually by starting with small hops before attempting higher jumps. This design not only prepares players for more intensive sections but can also serve as a warmup. In contrast to maximal jumps, small hops have a negligible risk of injury. One expert, who focuses on user groups with special needs, raised the concern that some players might not be ready for hops at all since the frequency someone uses jumps mainly depends on their exercise practice. Instead, the expert proposed starting with steps and balancing exercises before continuing with hops and larger jumps. As our goal was to design an engaging experience for everyone regardless of prior experience, we structured our game into four sequential levels, starting with tapping before increasing the intensity with hops, small jumps, and finally, maximal vertical jumps (see~\autoref{fig:game}).

The first three levels are structured similarly and tie the players' actions to the beats of a song. Such rhythmic movements are suggested by the literature~\cite{mueller2014} and have been used with great success in some of the most famous VR games, such as Beat Saber~\cite{beatsaber} or Ragnaröck~\cite{ragnarock}. Whereas all levels \rev{in our application} feature a different song, the length is always roughly two minutes with 128 beats per minute. Before starting the action, players receive a short introduction to every level and must perform each relevant movement correctly. During the level, a screen in the background displays the song progression and the players' performance, measured in successful hits and a derived score. Also, the game provides motivational feedback in the resting break between the levels.

To account for individual differences and fitness levels, we implemented three difficulty levels differing in the number and complexity of necessary movements. Starting at the lowest difficulty, the game automatically and unnoticeably switches to a higher level if players achieve a precision of at least 90\% or to a lower level if they miss more than 40\% of the last notes. With this design decision, we want to ensure that the game challenges all players without causing frustration. \rev{In particular, the game should ensure that players experience a feeling of success by performing at least half of the movements successfully. Conversely, it should raise the difficulty if players are not challenged and succeed in most interactions.} Ultimately, we aimed for an average hit rate between 60\% and 90\%. With this adaptive difficulty, we ensured that most players would achieve a high score boosting their motivation and confidence.

\subsubsection*{Level 1: Tap to the Beat}

In the first level \rev{(duration 01:54 min)}, yellow and purple tiles approach the players on four adjacent lanes. As the tiles reach the front line, players must tap on them with the correct foot --- yellow tiles require the left foot, whereas purple tiles map to the right foot. After tapping, players must retract their feet as they must not enter the playing field with both feet simultaneously. In the course of the song, the movement patterns become more complex and require players to switch feet quickly or tap crosswise. \rev{On the easiest difficulty, the game confronts players with 110 tiles, as opposed to 135 tiles on the hardest difficulty.} This level is mainly intended as a warmup and trains the players' lower body coordination, stability, and reaction time.

\subsubsection*{Level 2: Hopscotch}

The second level \rev{(duration 01:56 min)} advances on the first one by incorporating small hops. Similar to the child's game hopscotch, players have to hop with their left, right, or both feet on the correct tile. However, these hops are performed in place as the approaching tiles reach the players' position. The tiles move on five overlapping lanes, of which either one or two light up in yellow or purple to indicate the correct feet. Players must remain in the last pose between two hops, e.g., standing on one leg until another tile reaches them. Throughout our design phase, we learned that longer phases on one leg and fast switches between two and one leg are highly challenging and often do not fit the music. Instead, we mainly used slower two-legged jumping patterns or faster one-legged ``walking-style'' patterns. As before, the movement becomes more complex with time and incorporates the outer lanes more to force players to move from side to side. \rev{Depending on the difficulty level, players must perform between 108 and 129 hops.} This second level builds on the already trained balance and stability. Also, players must time their hop correctly to land on the tile when it lights up. This feature further trains coordination and neuromuscular control.

\subsubsection*{Level 3: Obstacle Course}

The last song-based level \rev{(duration: 02:08 min)} follows a different principle than the first two. This time, obstacles approach the players at every first beat of a bar. To gain points, players must avoid touching these impediments with their bodies. The most common obstacle is a low wall forcing players to jump at medium height. As our domain experts emphasized the importance of diversified training and dynamic exercises, we interleave the jump-over obstacles with lateral walls to both sides and barriers hanging from the ceiling. These force players to move sideways and duck down before jumping again. \rev{In total, this level features 65 obstacles, of which 24 are jump hurdles.} As personal size differences could pose an unfair disadvantage due to height differences, we scale the lower and upper walls according to the players' height during calibration. Also, the three difficulty levels affect only the obstacle size, not their frequency. Consequently, players must jump higher with increasing difficulty. This third level mainly focuses on muscle strength and endurance while preparing the players for the last level featuring maximal vertical jumps.

\subsection{Maximal Vertical Jumps}

After training the prerequisites for a good jump --- general fitness, balance, and coordination --- our last level focuses on teaching players a proper jump technique. In the beginning, players see an exemplary jump execution as part of a short introduction to the level. Next, players have to perform maximal vertical jumps and receive personalized feedback on their performance \rev{based on the four criteria below}. The game continuously records the players' movements. After detecting a jump, this data is analyzed with regards to four criteria of a safe and efficient jump that the domain experts mentioned:

\begin{enumerate}
    \item jump and land with both feet simultaneously
    \item land softly (forefeet touch the ground first) and absorb the impact with the entire body
    \item especially while landing, keep the knees in one line between feet and hips and do not cave them inward
    \item swing arms synchronously in a forward-upward arc until about chest height
\end{enumerate}

\rev{Analyzing most of these criteria, such as arm and leg synchronicity, is easily achieved by comparing positional differences to precalibrated thresholds. However, determining the players' landing style is more challenging since we do not track players' feet directly. Since the landing happens in only a split second, it is too fast for the precise infrared sync of the tracking system. Instead, the Vive trackers have to rely on their less-precise accelerometer. We use this tracking limits of the Vive system to our advantage; if players land hard without catching their impact through feet and legs, the measured position of the shin trackers descends for a brief time well below floor level. This vertical displacement directly correlates with the landing impact - a softer landing leads to less displacement and vice versa. So, our algorithm can use this tracking error as an indicator of the players' landing quality.} Furthermore, we implemented offset values until a violation of the above factors is considered insignificant to account for tracking imprecisions. After performing a jump, players see their performance in the four categories on one screen and receive a personalized message on another. This message summarizes their improvement from the previous jump, their worst-performing criterion in this jump, and practical instructions on how to improve on it in the next turn. Additionally, we visualize the last jump as a looping replay in front of the players, which allows them to study their movements. Multiple experts deemed combining extrinsic instructions with the opportunity for intrinsic feedback through jump visualizations highly valuable.

Apart from providing personalized instructions, we also calculate a jump score from the technical criteria and the effective jump height to reflect the performance. Together with the jump height, this value is listed on a highscore board, informing players of their improvements from jump to jump. In total, players perform five consecutive jumps and try to incorporate the feedback from the previous execution. After the last jump, the game ends. 

\section{Evaluation}

After designing our exergame \jump{}, we conducted an exploratory user study to explore how users perceive our novel training experience. Our primary research goal for our study was to confirm that our prototype provides an enjoyable user experience without causing unwellness or endangering players. Since players constantly remain in one spot through the experience, we are confident that the application is not likely to induce cybersickness. Additionally, we are interested in how the game's usability, appeal, and feedback contribute to the players' overall game experience.

Apart from these perceptional factors, we want to explore our exergame's motivational and physical effects on the players. Firstly, following our experts' feedback, we assume it improves the players' mood. Also, we hope that the players find the game's physical exercises challenging without frustrating or overly tiring them. Consequently, we are interested in how players perform in the various levels and where they see the future potential of such an exergame-based jump training.
\rev{Considering these motivations for our exploratory study, we employed various methods, including pre- and post-questionnaires, game performance data, and qualitative feedback, to answer our research questions:}

\noindent\begin{itemize}[leftmargin=*]\compresslist
    \item \rev{RQ1: How does our exergame affect cybersickness symptoms of players?}
    \item \rev{RQ2: How does our exergame affect the players' mood?}
    \item RQ3: How do players evaluate the player experience of our exergame and its usability as a training tool?
    \item RQ4: How do players perceive the physical activity and perform in our exergame?
\end{itemize}

\subsection{Pre-Post Questionnaires}
These questionnaires were administered before and after the gameplay. To answer \rev{RQ1}, we administered the simulator sickness questionnaire (SSQ)~\cite{kennedy1993simulator} in German~\cite{hosch2018simulator}. It measures three sub-categories, i.e., nausea, oculomotor disturbance, and disorientation, through 16 items on a 4-point scale. For \rev{RQ2}, we assessed the players' mood with the energetic and valence sub-categories of the German version of the Multidimensional Mood Questionnaire (MMQ)~\cite{wilhelm2007assessing}, consisting of four bipolar items on a 7-point Likert Scale (ranging from 0 to 6).

\subsection{Post-Questionnaires}
To measure the general game experience to answer RQ3, we used multiple sub-categories of the German version of the Player Experience Inventory~\cite{abeele2020development,graf2022pxi}: mastery, immersion, progress feedback, audiovisual appeal, challenge, ease of control, clarity of goals, and enjoyment (7-point Likert scale, ranging from -3 to 3). Additionally, we used the physical activity enjoyment scale (PACES-8)~\cite{mullen2011measuring} for RQ4, which includes eight bipolar items on a 7-point Likert scale (ranging from 1 to 7). The German item translation of PACES-8 was provided by one of the researchers, who used this scale in another research project.

\subsection{Game Performance Measures}
As we were interested in how players perform in our exergame to answer RQ4, we logged the necessary performance data for all participants. For each of the three rhythmic levels, we recorded the dynamic difficulty development and the hit ratio, i.e., how many steps, hops, and jumps over obstacles were successfully performed. Additionally, we collected the jump height and the performance in the four jump criteria for each jump in the final level.

\subsection{Qualitative Feedback}
We gathered qualitative feedback from the players through open-ended questions to understand their experience with \jump{}.
In these questions, we particularly focused on four topics: risks and benefits, safety and usability, long-term participation, and improvements.
We used the following open-ended questions to capture the perspectives of the players on these aspects:

\noindent\begin{itemize}[leftmargin=*]\compresslist
    \item \emph{``In your opinion, what are the risks and benefits of this VR game for you? Why?''}
    \item \emph{``Have you encountered any issues that affected your safety and usability during the VR gameplay? Why?''} %
    \item \emph{``If you could continue to use this VR game, how do you think that this would affect your long-term participation in jump training? Why?''}\footnote{\rev{This question is aimed at understanding participants' attitudes toward using this exergame in the long term, but we note that this cannot be taken as evidence that people would actually continue playing this game.}} %
    \item \emph{``Considering the VR game you played, what aspects would you like to change, and what aspects would you like to keep as they are? Why?''}
\end{itemize}

\section{Results}

\begin{table*}[t!]
\caption{The table shows the descriptive and the results of statistical tests of pre-post SSQ and MMQ (\textit{* indicates significance}).}
\resizebox{.975\textwidth}{!}{
    \centering
    \scriptsize
    \begin{tabular}{ccccccccc}
    \toprule
    \multirow{2}{*}{\textsc{Questionnaires}} &\multicolumn{2}{c}{\textsc{Pre-Que.}} &  \multicolumn{2}{c}{\textsc{Post-Que.}} &  \multicolumn{1}{c}{\textsc{Statistical}} & \multicolumn{1}{c}{\textsc{$P$}} &  \multicolumn{1}{c}{\textsc{Effect}} &
    \multicolumn{1}{c}{\textsc{Confidence}} \\
    & \emph{Mean} & \emph{SD} & \emph{Mean} & \emph{SD}  & \multicolumn{1}{c}{\textsc{Test}} & \multicolumn{1}{c}{\textsc{Value}} & \multicolumn{1}{c}{\textsc{Size}} & \multicolumn{1}{c}{\textsc{Interval (.95)}}\\
    \midrule
    
    \textsc{SSQ Total} & 15.41 & 13.41 & 21.09 & 14.76 & \textit{t(24)}=-1.71 & 0.100 & \textit{d} = -0.34 & [-12.533, 1.163] \\
    \textsc{SSQ-nau.}  & 15.26 & 11.02 & 33.58 & 17.44 & \textit{V}=4.5 & <0.001* & \textit{r} = -0.55 & [-28.620, -14.310] \\ 
\textsc{SSQ-ocu. dis.} & 15.77 & 16.80 & 11.82 & 14.19 & \textit{V}=72.5 & 0.220 & \textit{r} = -0.17 & [-7.580, 22.740]  \\
    \textsc{SSQ-dis.}  & 6.12  & 12.76 & 7.80  & 13.38 & \textit{V}=11 & 0.170 & \textit{r} = -0.19 & [-13.920, 13.920]\\

    \textsc{MMQ-ene.} & 3.6 & 1.35 & 4.1 & 1.61 & \textit{t(24)}=-1.50 & 0.146 & \textit{d} = -0.30 & [-1.187, 0.187]  \\ 
    \textsc{MMQ-val.} & 4.6 & 1.24 & 4.2 & 1.55 & \textit{V}=112 & 0.092 & \textit{r} = -0.24 & [-0.000, 1.250]\\
    \bottomrule
    \end{tabular}}
    \label{tab:gameperf_prepost}
    \Description{The table shows the descriptive values and the results of statistical tests of SSQ and MMQ questionnaires.}
\end{table*}

\begin{table*}[t!]
\caption{The table shows the descriptive values of the PXI and PACES-8 questionnaires: The player experience and physical activity enjoyment of players were generally high.}
\resizebox{\textwidth}{!}{
    \centering
    \small
    \begin{tabular}{cccccccccccccccccccc}
    \toprule
    
    \multicolumn{2}{c}{\textsc{PXI-enj.}} & \multicolumn{2}{c}{\textsc{PXI-mas.}} &  \multicolumn{2}{c}{\textsc{PXI-imm.}} &  \multicolumn{2}{c}{\textsc{PXI-pro. fee.}} & \multicolumn{2}{c}{\textsc{PXI-aud. app.}} & 
    \multicolumn{2}{c}{\textsc{PXI-cha.}} &  \multicolumn{2}{c}{\textsc{PXI-eas.}}  &\multicolumn{2}{c}{\textsc{PXI-cla.}} & \multicolumn{2}{c}{\textsc{Paces-8}} \\
    \emph{Mean} & \emph{SD} & \emph{Mean} & \emph{SD} & \emph{Mean} & \emph{SD} &\emph{Mean} & \emph{SD}  & \emph{Mean} & \emph{SD} & \emph{Mean} & \emph{SD} & \emph{Mean} & \emph{SD} &\emph{Mean} & \emph{SD} &\emph{Mean} & \emph{SD} \\
    \midrule
    1.95 & 0.79 & 0.93 & 1.19 & 2.03 &  0.75 & 1.6 & 0.88 & 1.33 & 0.98 & 1.71 & 0.80 & 1.99 & 0.84 & 2.2 & 0.93 & 5.44 & 0.91 \\
    \bottomrule
    
    \end{tabular}}
    \label{tab:gameperf_px}
    \Description{The table presents the descriptive values of the PACES-8 questionnaire and the PXI sub-categories. The player experience and physical activity enjoyment of players were generally high.}
\end{table*}

Twenty five participants (15 female, 10 male, \mean 24, \sd 6.01 years) were recruited for our study. Twenty of them had prior VR experience. However, only two reported using VR frequently (1-2 times per month). Of the rest, twelve participants rarely used VR devices (1-2 times per year), and the remaining had only one to two prior sessions. Asked for their exercising habits, only two participants stated not to be exercising. Among the rest, ten participants exercised one or two times per month at most. Eleven reported exercising at least once per week, and two trained daily. Additionally, the participants generally rated physical exercise enjoyment slightly positive (\mean 1.24, \sd 1.45, range -3 to 3).

After learning about our research objectives and signing informed consent, participants completed the first part of the questionnaire, assessing demographics, pre-SSQ, and mood. Upon completion, we introduced the participants to the Vive Pro VR headset~\cite{vive} and assisted them in attaching the Vive Trackers~\cite{viveTracker} to their shins and waist. After starting the game, the participants received an introduction to the controls and calibrated their avatar before playing the four levels. After the playthrough, they completed the post experience questionnaires and answered the open-ended questions. The duration of the study was 35 minutes on average.

\begin{figure}[t]
    \centering
    \includegraphics[width=\columnwidth]{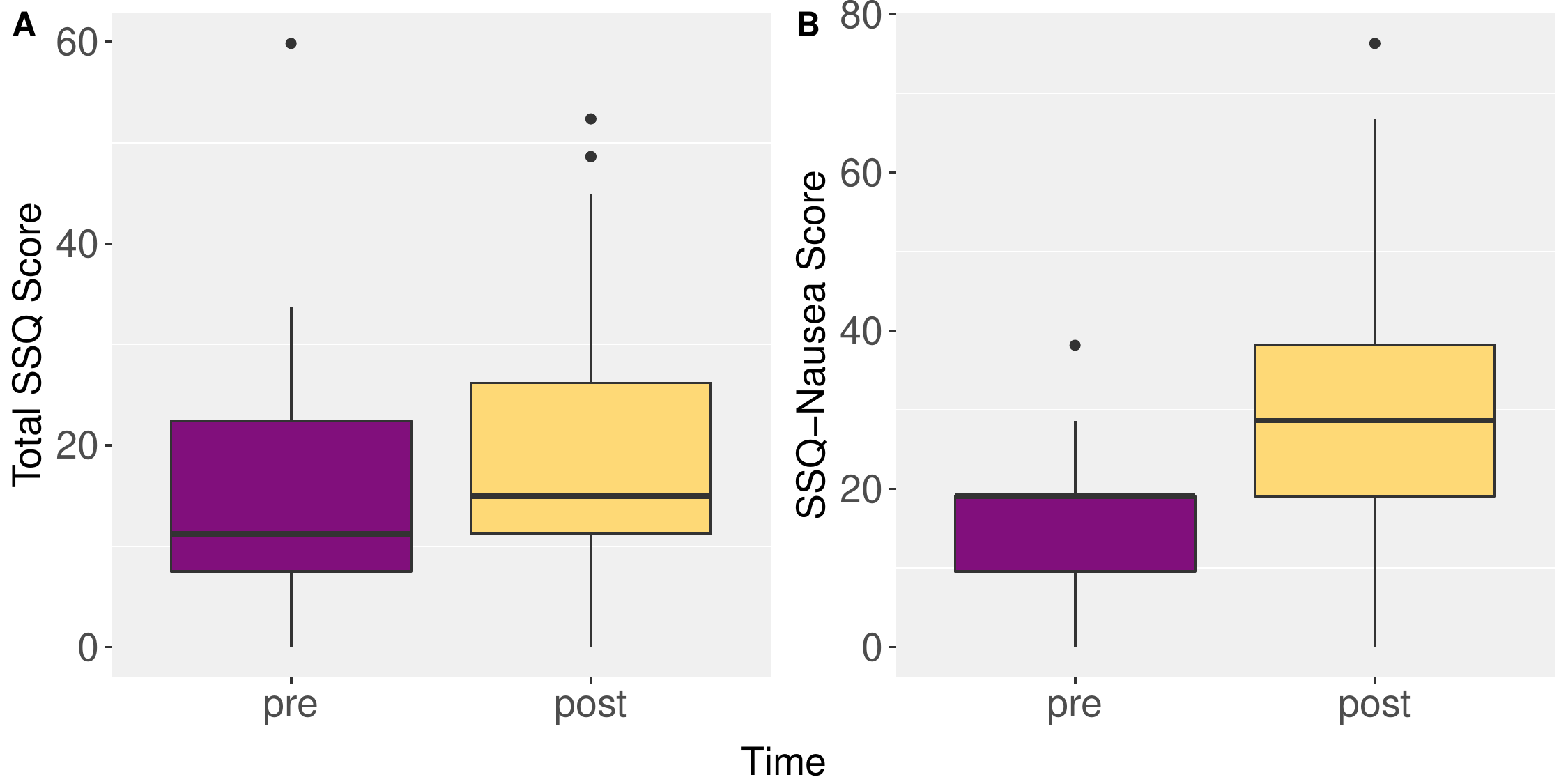}
    \caption{Although the participants' Total SSQ ratings were not significantly different between pre- and post-time points, post-SSQ-Nausea scores were significantly higher compared to the pre-SSQ-Nausea scores.}
    \Description{The figure shows the boxplots of pre- and post-scores of Total SSQ and SSQ-Nausea.}
    \label{fig:boxplot}
\end{figure}

\subsection{Pre-Post Questionnaires}
We conducted Shapiro-Wilk tests to check the normality assumption of the data.
When the data was normally distributed, we used paired t-tests and reported the effect size with Cohen's \textit{d}.
In the case of non-normally distributed data, we used Wilcoxon-signed rank tests and reported \textit{r}, Pearson’s correlation coefficient, as the effect size measure.
We followed \citet{Cohen92quantitativemethods}'s recommendations to intepret these effect sizes.
\autoref{tab:gameperf_prepost} lists the results of statistical tests as well as descriptives of each questionnaire and their sub-categories.

A wilcoxon-signed rank test indicated that participant's post-nausea scores were significantly higher than pre-scores, \textit{V}=4.5, \psmall, \autoref{fig:boxplot}b. However, Total SSQ scores and the other sub-categories of SSQ did not indicate significant differences between pre- and post-values.

Due to the nature of exercise, people tend to sweat. Whereas sweating can be a symptom of cybersickness, in our case, it is more likely an effect of the physical effort during training~\cite{armstrong1998effects}. Therefore, we also performed the SSQ analysis while excluding the sweating item. Since this item is considered only in the calculation of Total SSQ and SSQ-Nausea categories, we report only their analysis. Total SSQ scores were not significantly different between pre- (\mean 12.42, \sd 13.30) and post-time points (\mean 12.42, \sd 13.73), \textit{t(24)}=0, \textit{p}=1, $d$=0. Similarly, the players' nausea ratings did not significantly differ between pre- (\mean 7.63, \sd 9.54) and post-measurements (\mean 11.45, \sd 13.49), \textit{t(24)}=-1.29, \textit{p}=0.211, $d$=-0.26.

Neither the energetic nor the valence sub-scale of the MMQ showed significant differences between before and after scores of playing the game.

\subsection{Post-Questionnaires}
We report the descriptive values of the questionnaires administered as only post-game measures in \autoref{tab:gameperf_px}. The findings indicate that all measures were rated with a positive tendency.
Whereas most constructs highlight a highly positive experience, the PXI-mastery sub-category indicates that the players did not feel a particularly high mastery in the game.

\subsection{Game Performance}
For the first three levels, we logged the participants' hit ratio as the main performance measure. By adding the dynamic difficulty adaption in our game design process, we aimed for an overall success rate between 60\% and 90\%. This goal was achieved for all three levels \textit{Tap to the Beat} (\mean 78.4\%, \sd 15.8), \textit{Hopscotch} (\mean 69.6\%, \sd 15.4), and \textit{Obstacle Course} (\mean 85.1\%, \sd 8.6). On average, participants spent 41.77\% of the time in the medium difficulty (\sd 23.63), followed by the hardest difficulty (\mean 35.33\%, \sd 25.11) and the easiest difficulty (\mean 22.88\%, \sd 19.33).

The participants mainly did not improve their jump height in the final level over the five jumps. The average height remained roughly the same from 38.28cm (\sd 9.12) in the first round to 37.66cm (\sd 9.10) in the last execution. However, participants improved their technique score slightly over the five jumps, starting from an average rating of 71.45\% (\sd 17.18) and ending at 75.45\% (\sd 15.30). 

\rev{Lastly, we also logged how often the game provided feedback to the players. Although this data is purely descriptive, it gives a good impression of how our system responded to the players' movements.} Generally, the game advised participants for 62\% of the jumps to pay attention to the correct knee movement during the landing. This criterion was followed by a too-hard landing, an issue in 26\% of the cases. Lastly, the arm swing was too high for 20\% of the jumps and not fully parallel for 8\% of the jumps. In contrast, asynchronous feet movements during the lift-off and landing were not an issue.

\subsection{Qualitative Feedback}
We collected qualitative data in the form of open-ended questions. Before the analysis, the data was translated into English using DeepL Pro~\cite{deepl} and was checked by one of the first authors for inaccurries. After this, we used Dovetail tool to code the data~\cite{dovetail}. 

One of the first authors analyzed this data using a reflexive thematic analysis approach~\cite{braun2021can,braun2019reflecting}. Before the analysis, the author decided on four deductive categories: \emph{risks and benefits}, \emph{safety and usability}, \emph{participation}, and \emph{improvements}. The author coded the interview data using inductive codes (e.g., ``injury risk'', ``accessibility of exercise opportunities'', ``standing zone should be improved'') under these categories. Following this step, they performed an affinity mapping activity and based on this, they created following themes.

\subsubsection*{\textbf{Theme 1: The main advantages of JumpExTra VR are accessibility and enjoyment, however, participants also reported injury concerns.}}
The participants mainly attributed the advantages of this game to two factors. The first factor is the accessibility of physical exercise opportunities: \quoting[P4]{[...] you don‘t need to go to the gym since you can easily exercise at home}. Secondly, they emphasized the enjoyment aspect of \jump{}: \quoting[P9]{More fun while exercising}. Interestingly, one participant reported both the pros and cons of immersion in this game: \quoting[P6]{Forgetting the real world is a disadvantage and being completely immersed in the world is an advantage}. Another player highlighted the positive side of this game by comparing it to another commercial alternative: \quoting[P16]{More movement for players than in other VR games like Beat Saber}.

\jump{} was found to be associated with some drawbacks. Some players reported the possibility of losing physical-world awareness while playing this game: \quoting[P4]{A disadvantage would be that you might get too "infatuated" with the virtual world and neglect real life}. Additionally, many participants were concerned about the potential injury risks. These were attributed to various causes, but mostly to falling: \quoting[P14]{I found it risky to fall down}. 
The risk of physical collisions were also pointed out: \quoting[P8]{You might bump into objects in the real environment}. Referring to the technical feedback in the last level, one participant stressed the need for similar training instructions for the more gameful levels, too, \quoting[P3]{because, if [they] do the tasks wrong, injuries could follow}. Keeping the balance was mentioned as an issue for a few: \quoting[P7]{It's kinda hard to keep balance sometimes, the risk is that you can fall on the ground}.

\subsubsection*{\textbf{Theme 2: Overall JumpExTra VR did not cause serious safety and usability issues. Yet, there were some instances reported by players.}}
Even though participants had no severe issues regarding safety and usability, a few had difficulties understanding where they were physically located: \quoting[P2]{I couldn't remember where I was in the room}.
Some echoed the safety issues related to having accident, falling down, and losing balance. 
The difficulty of staying in the standing zone of \jump{} was also noted: \quoting[P5]{I had trouble staying within the designated play area. I got too far ahead in places, and so I couldn't get the triggers to work properly on the first play}. 
Notably, the players reported on some usability problems due to the technical VR setup: first, \quoting[P23]{Problems with sharp vision, which strained the eyes a lot}. Second, \quoting[P16]{Tracking did not work perfectly}.
Third, \quoting[P8]{The cable on [their] back was disturbing}.
Nevertheless, only one participant encountered issues with the leg trackers not staying in position, and another participant had experienced usability problem due to color blindness.

\subsubsection*{\textbf{Theme 3: For most players, JumpExTra VR would positively impact participation in jump training.}}

Most participants agreed that the chance of using \jump{} would positively affect their long-term participation in jump training. Some commented that \quoting[P6]{it would be good for [their] fitness}.
According to one player, especially over time, one would get better at performing jumps and at the evaluated aspects, which \quoting[P3]{increases the average jumping power and also reduces the risk of injury}.
The feedback feature of the game was particularly appreciated by some participants and mentioned as a reason for long-term use: \quoting[P10]{My jump will probably improve through feedback}.
For a few participants, the motivating or fun nature of \jump{} played a role: \quoting[P19]{The design and the atmosphere is very pleasant. It would motivate me}.
In particular, one player emphasized the advantage of using games for physical exercises: \quoting[P24]{[...] during the games you don't really notice how you practice your jumps and therefore it is not so monotonous}.

However, some participants would not continue using \jump{}. For a few, this was due to safety and comfort issues that can occur when jumping in VR: \quoting[P5]{I didn't feel safe enough to jump with full power. The headset wobbled too much for that and I was afraid of not landing properly or damaging the headset.}. 
Tracking imprecision was also pointed out as a reason: \quoting[P25]{I also found the jump training segment to be too inaccurate from a tracking standpoint}. 
Additionally, there were other reasons reported by the players, such as a general dislike for VR games, loss of interest after a while, or difficulties in maintaining habits.

\subsubsection*{\textbf{Theme 4: Players suggest improvements for JumpExTra VR about the game world, feedback, standing zone, and varition of game levels.}}

Overall, we received a lot of feedback for our game design. A few participants would even \quoting[P13]{change nothing}. Others emphasized the parts they liked the most, like \quoting[P6]{[...] definitely keep the music and tutorials}. Still, we received many ideas of how to improve \jump{}. Some players proposed improvements to the game world, such as \quoting[P8]{mak[ing] the game environment more colorful, beautiful, lively}, and \quoting[P14]{a more accurate virtual body would be desirable to play better}. 
A few wished to have more feedback: \quoting[P20]{What I would change is the feedback. A sound or buzzer was missing if the jump or obstacle evasion were successful or not.}.
A part of the players did not like constantly looking down in the gameplay and improving the standing zone was recommended: \quoting[P3]{I wished for an alert once you are outside the designated play zone}.
More variety was also suggested for jump training exercises: \quoting[P10]{Combine the three levels for more variety (tap, jump, dodge)}.

\section{Discussion}
The main goal of this research project was to provide playful and safe VR-based jump exergame. Therefore, we considered both prior literature and the findings of expert interviews to design \jump{}. As the final step, we evaluated our prototype in a mixed-methods user study ($N$=25).
In the following section, we discuss the findings by focusing on the hypothesis and research questions and provide design implications that can help researchers and practitioners to expand on full-body exertion experiences.

\subsection{RQ1: How does our exergame affect cybersickness symptoms of players?}
\rev{Our results show that the players reported higher SSQ-nausea scores after playing the game.} This finding supports previous research~\cite{szpak2020exergaming} showing that VR exergames can cause symptoms of cybersickness.
However, we did not observe increased ratings for the remaining categories of SSQ: oculomotor disturbance, disorientation, and the total score.
We suspect the increased sickness scores were due to the sweating item of SSQ. The effect of physical training on sweating has been shown in the literature~\cite{armstrong1998effects}. Our additional SSQ analysis, excluding the sweating item, supports this assumption; the results show no significant effect between pre- and post-measurements. As suggested by~\cite{szpak2020exergaming}, we also emphasize the potential overlap between cybersickness and physical activity symptoms. Our results underline that high-intensity full-body exercises can be safely performed in VR without risking discomfort.

\subsection{RQ2: How does our exergame affect the players' mood?}
\rev{Feedback from our domain experts and literature~\cite{chan2019therapeutic} suggests that the physical activity in our exergame is likely to improve the players' mood. To explore this potential effect,} we assessed the players' energetic arousal and valence. \rev{Our results from the MMQ show that neither of the two subscales revealed any significant difference.} We explain this outcome two-fold. In case of the energetic arousal subcategory, most players stated being physically exhausted and sweaty after playing the game. They explicitly attributed their state to two external reasons: summer heat and headset fit. Unfortunately, the study coincided with an extreme summer drought with high temperatures. Also, COVID-19 measures forced us to use easily cleanable headset foams which accumulated extra heat. Both factors added to the expected exhaustion and potentially caused players to feel tired rather than stimulated. Still, as players generally appreciated the physical challenge, we are confident that such full-body exergames can benefit the players' physical and mental well-being in the long run.

In contrast, the slight decrease in valence scores might be connected to the low scores on the PXI's mastery dimension, indicating that participants did not feel particularly good at playing the game. Due to the time constraints of the study, we could let participants try the game only once. However, such fast and timed movements are always hard when attempting them for the first time. Therefore, we speculate that players would feel more relaxed and capable with further repetitions. In turn, a feeling of success that was missing in this first run might also positively impact participants' valence.

\subsection{RQ3: How do players evaluate the player experience of our exergame and its usability as a training tool?}
\jump{} was implemented based on prior literature and expert interviews. Therefore, we see the high ratings of PXI as a result of the detailed design process of our game. Aligning with previous studies~\cite{finkelstein2010astroautism,ioannou2019virtual,lehtonen2019movement}, we show that jump-based exergames can lead to positive player experiences. With this, we extend the results of these studies into the training realm of jumping in VR.

Overall, the players highly enjoyed playing this game and found it immersive and appealing. The high PXI progress feedback scores show that the game provided comprehensive feedback regarding players' progress. This finding is further supported with qualitative data as the players appreciated the game's feedback feature: \quoting[P10]{feedback will probably improve my jump}.
We believe that the instructions given before each level (clarity of goals), interactive tutorials (ease of control), and dynamic difficulty (challenge) are reflected in the high scores in the respective PXI sub-categories.
However, players criticized the need to look down at their feet constantly. This pose is uncomfortable and could potentially lead to postural degradation. Instead, visual and audio feedback informing players upon a hit or miss could reduce the urge to check on their own movements.

Even though most players stated they would continue using \jump{}, \rev{we cannot take this feedback as evidence for long-term retention. Instead, we emphasize the need for longer, focussed studies. Furthermore, players} requested visual improvements, namely a more lively world and a better-matching avatar. Also, they proposed combining the levels for more variety in gameplay. Especially this last point is vital for the long-term success of future full-body exergames. To secure retention, developers should focus on varied gameplay with high replayability. Rhythmic elements used in our game or famous titles like Beat Saber can be easily extended by adding new songs and mappings. The repetitive technical jump training in our last level bears a much greater challenge and will likely suffer from decreasing interest. Hence, it remains an open research question on how to enhance retention for advanced training routines as well.

\subsection{RQ4: How do players perceive the physical activity and perform in our exergame?}

Overall, all participants performed well in our exergame. In particular, our results reveal that the individual success rates for all levels are well within our anticipated window of 60\% to 90\%. Hence, we can assume that our dynamic difficulty adaptation worked well in aligning the exergame's challenge to the players' individual abilities. Within our sample, the participants were also able to increase their personal technique score in the final level by 5.60 \% on average. In contrast, we did not see any improvements in jump height. These observations match the domain experts' feedback that muscular changes take multiple weeks, whereas technical improvements are quickly realizable. \rev{However, we note that our results are purely descriptive. Determining the system's accuracy would require an additional evaluation in a professional movement analysis lab.} In terms of perceptual effects, the high values in the challenge sub-category of the PXI show that the participants were positively challenged without feeling overly taxed. Additionally, the PACES-8 results highlight that the participants enjoyed the physical activity in \jump{}. Lastly, we observed improvements in the players' skills. Initially, many participants had immense difficulties with coordination, used the wrong foot, or could not follow the song's rhythm. As their performance improved throughout the course of the game, we believe that further sessions would positively influence players' body control and neuromuscular coordination.

\subsection{Design Implications}

\subsubsection*{Avoid the Unintended Forward Drift (UFD)}
In our study, the frequent tapping and jumping on the spot led to small, unnoticed forward movements that required the players to monitor their position to prevent leaving the play area. This effect is similar to the unintended positional drift (UPD)~\cite{nilsson2013unintended} that is observed for walking-in-place locomotion techniques. In an analogy, we name our observed effect unintended forward drift (UFD). Unfortunately, we do not see an easy solution except for integrating a warning and pausing the game when players leave the designated area \rev{(matches with guidelines of \citet{xu2020results})}. Forcing players to remain precisely in one spot would increase the necessary mental effort and potentially spoil the game experience. An option would be to integrate omnidirectional gameplay where players jump in all directions and not just forward. However, such game design is challenging because it adds events behind the players' backs.

\subsubsection*{Consider Technical Constraints}
\rev{Currently available VR hardware was not designed for tracking explosive full-body movements. Its limitations manifest in the need for additional hardware (e.g., Vive trackers) and the likeliness of tracking errors. Although these technical limitations limit the practical applicability for today's commercial games, we are confident that they are resolvable with future generations of VR headsets. Instead, we share the lessons we have learned by using the current state of technology to achieve a jump-centered exergame.}
In our pilot tests, we experimented with different approaches to attaching the trackers to the players' legs. In the end, we decided on the shins. Compared to typical shoe adapters, trackers are more stable, more comfortable to wear, and easier to attach in this position. However, during the study, we sometimes noticed that trackers slipped slightly, especially if participants wore slick pants. Even if trackers remain in place, Vive lighthouse tracking is not accurate enough for fast leg movements despite being one of the most precise VR systems. Consequently, we noticed considerable jittering during the game. This problem is not severe in the rhythmic levels and was not noticed. However, when using the tracking data to display a replay of the vertical jump, users see this suboptimal tracking quality. Consequently, we recommend applying a slight low-pass filter to remove artifacts in cases where players see their own movements, like in our replay. 

Besides visual artifacts, the tracking-related inaccuracies also influence the quality of the jump-technique feedback. Paired with a variety of body types and postures, these technical constraints can increase the false-positive rate of mistake detections, such as problematic knee movement. To avoid wrong corrections manifesting in the players' jump technique, automated feedback should be a tool only to support their own critical reflection on their movements. In this context, we noticed that players often did not recall what the optimal jump, shown in the tutorial, looked like and could not draw proper conclusions for their jump. Hence, we recommend tying visualizations of the ``optimal performance'' with the replay and complementing both with the less-precise automatic feedback. This combination further improves the players' critical view of their performance and may lead to lasting technical improvements.

\subsubsection*{Account for Safety Concerns in Fully-Body VR Exergames}

\rev{We extend the work of~\citet{mueller2014} by providing safety implications specifically for full-body VR exergames.}
As we have seen with the UFD, performing full-body movements in VR has safety concerns that must be accounted for early in the game design pipeline. In particular, one of the main reasons we initially chose to focus on the vertical jump is its compatibility with normal-sized tracking areas. Other movements, such as forward jumps, transport users quickly to the play space's borders or beyond. This consideration is particularly important for game developers who must account for the varying consumer play spaces that often do not have the generous size of a $16m^2$ VR lab. 

Another critical concern is the potential risk when using a cable-bound VR headset, such as the Vive Pro. In our case, players mostly did not move enough in the tracking area to encounter cable-related issues. However, one of the authors was always present to monitor the participants' behavior. Of course, this approach is no solution for commercial applications. Lastly, we noticed that participants generally had more problems with balance than in real life. Since accidental tripping in smaller tracking areas could easily lead to collisions, developers should consider this observation.

\subsubsection*{Incorporate Rhythmic Elements with Fluent Movement Patterns}

Similar to other successful exergames like Beat Saber, we aligned the players' movements in the first three levels to the beats of a song. This game design was widely appreciated by the participants who asked us to \quoting[P6]{[...] definitely keep the music and tutorials}. However, it is essential to limit the length of each level, as fast jumping and hopping exercises are exhausting. In our case, we used only two-minute songs. Also, we incorporated frequent ``resting passages'' featuring fewer and slower interactions to allow players to regain their breath.

In our game design process, we experimented with various movement patterns and learned which worked well or harmed the game experience. First, frequent switches between single-leg and double-leg movements are highly challenging and require excellent coordination and balance. Also, the rhythm of steps and jumps should be mostly regular and seldom change. For instance, patterns on every, every second, or every fourth beat work perfectly fine. However, repeated transitions between these intervals interrupt the players' flow and cause frustration due to missed notes. For single-leg tapping and hopping, we found that a ``walking-like'' pattern, i.e., switching the feet between every note, works well. A good equivalent for two-legged moves is repeatedly jumping into a narrow and a wide stance.

\subsubsection*{Adapt the Difficulty According to Players' Fitness Level and Improvements}

\rev{Similar to prior exergame studies~\cite{xu2020results,mueller2016}, we emphasize the importance of using a dynamic difficulty adaptation---a key feature in ensuring a satisfying game experience for all players. Our game} constantly monitors the players' performance and switches to another difficulty level on demand. This functionality allows us to challenge every player regardless of their fitness level and ability without overtaxing beginners. Our analysis of the game's performance measures revealed that the participants spent most of the time in the medium and hard difficulties. In contrast, the easiest difficulty was rarely used. Hence, we see potential in better balancing the difficulty levels and improving the transition parameters. However, estimating the effective difficulty of a particular song mapping is not easy. For instance, we explained in the previous design implication that fewer interactions do not automatically translate into an easy gameplay.
\rev{In general, our three difficulty levels provided an optimal challenge for most players. However, our logged performance metrics revealed that few players still underperformed with the easiest difficulty, whereas others were not even challenged by the hardest level.} Hence, we recommend adding more difficulty levels to fit every fitness level. %

\subsection{Limitations}

We used a Vive Pro headset with Vive Trackers in our study. This decision was mainly motivated by the improved tracking quality compared to alternatives, such as the Kinect. However, the use of a cable-bound system also led to usability and safety concerns. Whereas we are confident that active monitoring during the exergame prevented any dangerous situations, we accept that participants might have felt more insecure or disturbed by the VR system. \rev{Additionally, we only conducted an exploratory user study to understand how players experience our exergame. Consequently, we cannot make conclusive statements about the accuracy of the automated feedback suggestions in the last level.} \rev{Overall, we also note that the use of additional hardware limits access to this exergame, and we hope that future technological advancements will open up new possibilities for tracking full-body movements in VR.}

For the qualitative data in our research, we reflect on our background and potential research interests that might have impacted the analyzing process~\rev{\cite{newton2012no, braun2021can}}. One of the first authors involved in the expert interview analysis has a computer science background and has published on VR, locomotion, and games research before. The other first author, who conducted both qualitative data analyses, has research experience in VR and exergames for varying user groups and a background in psychology and cognitive systems. Hence, they may have introduced bias into the analysis due to their interests and background. However, we also consider that the combination of these different specialties and perspectives enriches the data processing step.

\section{Conclusion}

VR exergames can provide a great motivation to pursue a more active lifestyle and exercise regularly at home. However, most available games \rev{track and focus primarily only hand movements in their routines, which leaves the lower body, already weakened by all-day sitting, severely undertrained and untargeted.} In this work, we explored the potential of full-body VR exergames using the example of vertical jump training. Therefore, we interviewed nine domain experts and combined their feedback with insights from prior research into our exergame prototype \jump{}. In the first three levels, the game trains lower body coordination, stability, and endurance, before providing technical feedback on the execution of maximal vertical jumps.

Additionally, we conducted an exploratory user study to evaluate how players perceive the training experience with \jump{}. Our results reveal that the participants appreciated the physical challenge and enjoyed our jump-centered exergame. Based on the participants' feedback, we provided a set of design implications that can guide future work on full-body VR exergames and help developers design engaging experiences. In future work, we want to extend our research by evaluating the long-term effects of our exergame and compare the training effects of this game with supervised training.

\begin{acks}
This work was supported by the German Federal Ministry of Education and Research (BMBF).
\end{acks}

\bibliographystyle{ACM-Reference-Format}
\bibliography{ref.bib}

\balance
\end{document}